\documentclass[preprint,prd,showpacs,preprintnumbers,amsmath,amssymb,eqsecnum,nofootinbib]{revtex4}

\usepackage{graphicx}
\usepackage{dcolumn}
\usepackage{bm}

\newcommand{\be}{\begin{equation}}
\newcommand{\ee}{\end{equation}}
\newcommand{\bea}{\begin{eqnarray}}
\newcommand{\eea}{\end{eqnarray}}
\newcommand{\bi}{\begin{itemize}}
\newcommand{\ei}{\end{itemize}}
\newcommand{\benu}{\begin{enumerate}}
\newcommand{\eenu}{\end{enumerate}}
\newcommand{\nn}{\nonumber}
\newcommand{\tbeta}{\widetilde{\beta}}
\newfont{\bg}{cmr10 scaled\magstep4}

\begin{document}

\preprint{KEK-TH 1487,\quad KUNS-2357,\quad MISC-2011-17}

\title{Squark Contributions to Photon Structure Functions\\
and Positivity Constraints
}

\author{Yoshio Kitadono}
 \email{kitadono@phys.sinica.edu.tw}
 \affiliation{
 High Energy Accelerator Research Organization (KEK),
 1-1 Oh-ho, Tsukuba, Ibaraki 305-0801, Japan\\
Current address: Institute of Physics, Academia Sinica, Taipei, Taiwan}

\author{Ryo Sahara}
\email{sahara@scphys.kyoto-u.ac.jp}
\affiliation{Dept. of Physics, Graduate School of Science, Kyoto University,
 Kitashirakawa, Kyoto 606-8502, Japan}

\author{Tsuneo Uematsu}
\email{uematsu@scphys.kyoto-u.ac.jp}
\affiliation{Dept. of Physics, Graduate School of Science, Kyoto University,
Kitashirakawa, Kyoto 606-8502, Japan}

\author{Yutaka Yoshida}
 \email{yyoshida@post.kek.jp}
 \affiliation{
 High Energy Accelerator Research Organization (KEK),
 1-1 Oh-ho, Tsukuba, Ibaraki 305-0801, Japan
\vspace{0.1cm}}%

\date{\today}

\begin{abstract}
Photon structure functions in supersymmetric QCD are investigated
in terms of the parton model where squark contributions are evaluated.
We calculate the eight virtual 
photon structure functions by taking the discontinuity
of the squark massive one-loop diagrams of the photon-photon forward amplitude.
The model-independent positivity constraints derived from
the Cauchy-Schwarz inequalities are satisfied by the squark parton
model calculation and actually the two equality relations hold for
the squark contribution. We also show that our polarized photon 
structure function $g_1^\gamma$ for the real photon leads to the
vanishing 1st moment sum rule, and the constraint $|g_1^\gamma|
\leq F_1^\gamma$ is satisfied by the real photon. We also
discuss a squark signature in the structure function $W_{TT}^\tau$.
\end{abstract}

\maketitle

\section{Introduction}

The Large Hadron Collider (LHC) \cite{LHC} has restarted its operation
and it is anticipated that the signals for the Higgs boson as well as 
the new physics beyond Standard Model, such as an evidence for the 
supersymmetry (SUSY), might
be discovered. Once these signals are observed more precise measurement needs 
to be carried out at the future $e^{+}e^{-}$ collider, so called 
International Linear Collider (ILC) \cite{ILC}.  

It is well known that, in $e^+ e^-$ collision experiments, the cross section
for the two-photon processes $e^+ e^- \rightarrow e^+ e^- + {\rm hadrons}$
dominates at high energies over 
the one-photon annihilation process $e^+ e^- \rightarrow \gamma^*
\rightarrow {\rm hadrons}$. We consider here the two-photon processes in the
double-tag events where both of the outgoing $e^+$ and $e^-$ are detected.
Especially, the case in which one of the virtual photon
is far off-shell (large $Q^2\equiv -q^2$), while the other is close to
the mass-shell (small $P^2=-p^2$), can be viewed as a
deep-inelastic scattering where the target is a photon rather than a 
nucleon~\cite{WalshBKT}. In this deep-inelastic scattering off photon targets, 
we can study the photon structure functions \cite{Review}, 
which are the analogues of 
the nucleon structure functions.

In order to analyze the two-photon process including new heavy particles 
at ILC, 
it is important to consider the mass effects of
the new heavy particles, like supersymmetric particles.
In this paper we investigate contribution from the squarks, the super-partner
of the quarks, to the photon structure functions. Before the supersymmetric
QCD radiative effects are studied taking into account the mass effects, 
it is worthwhile, first, to investigate squark contributions
to the photon structure functions through the pure QED interaction fully
taking into account the squark mass effects.
We evaluate the eight virtual 
photon structure functions by taking the discontinuity
of the squark one-loop diagrams of the photon-photon forward amplitude.
We study the model-independent positivity constraints whether squark parton
model calculation satisfies these constraints.
The real photon case is recovered by putting $P^2$ (target photon mass
squared) equal to zero.

The real unpolarized photon structure functions, 
$F_{2}^{\gamma}$ and 
$F^{\gamma}_{L}$ were investigated by the parton model (PM) 
in \cite{QPM} and were studied by the operator product 
expansion (OPE) supplemented with the renormalization group equation 
method \cite{Witten,BB} and were calculated by improved PM powered by 
the evolution equations \cite{DJSWW,GR,MVV,VMV}.
In the case that the mass squared of the target photon is non-vanishing ($P^2
\neq 0$), we can investigate the virtual photon structure functions. 
The unpolarized virtual photon structure functions were studied to LO
in \cite{UW1} and to NLO in \cite{UW2,Rossi, Borzumati-Schuler,Chyla}. 
Parton contents were studied
in \cite{Drees-Godbole,GRStrat} and the target mass effect of virtual photon
structure functions in LO was discussed in \cite{MathewRavi}. 
The heavy-quark mass effects in photon structure functions were
studied in the literature \cite{GR,GRS1,GRS2,GRStrat,Fontannaz,AFG1,AFG2,GRSchi,CJKL,CJK}.
See, for example, the recent work by pQCD \cite{USU,KSUU1,KSUU2,Kitadono}, by 
AdS/QCD \cite{Yoshida} and references therein. 
The polarized photon structure function $g^{\gamma}_{1}$ was investigated 
with pQCD up to the leading order (LO) \cite{Sasaki1,Sasaki2}, and the next-to-leading order (NLO) \cite{SU,SV,GRS1,GRS2}. 

The general forward photon-photon scattering amplitude is characterized by the helicity amplitudes and those are decomposed into eight tensor structures 
\cite{BCG,BGMS,BM,CT}. But we have four-independent structure functions in the case that the target photon is on shell. 
The results of four-independent real photon structure functions $W_{TT}$, $W^{a}_{TT}$, $W^{\tau}_{TT}$, $W_{LT}$
by the Quark Parton Model (QPM) to the Leading Order (LO) in QED were derived in Ref. \cite{SSU1} and the results of eight-independent virtual photon structure functions by the QPM were obtained in Ref. \cite{SSU2} (also see Ref. \cite{Schienbein}). In these references \cite{SSU1,SSU2}, the three positivity constraints were derived for the virtual photon target by using the Cauchy-Schwarz inequality and those reduce to one constraint in the real photon limit. All results satisfied with these constraints up to the leading order in QED.

On the other hand, 
the photon structure functions in supersymmetric theories were studied in 
Refs. \cite{Reya,SS,DGR,RW} 
up to the leading order in SUSY QED. 
In these references, the real photon structure functions $F_{2}^{\gamma}$ 
and $F_{L}^{\gamma}$ were considered instead of the four-independent photon 
structure functions. 
Furthermore the study of the polarized real photon $g^{\gamma}_{1}$ will be 
important theoretically and phenomenologically,
since the polarized photon structure functions $g^{\gamma}_{1}$ has a 
remarkable sum rule, $\int_{0}^{1} g^{\gamma}_{1}(x,Q^2)dx=0$ 
\cite{Bass,ET,NSV,FS,BBS}. 
The another constraint $|g^{\gamma}_{1}|\leq F^{\gamma}_{1}$ is derived in 
Ref.\cite{SV,GRS1}. 
We will show that our result for the polarized photon structure function 
satisfies this sum rule and the constraint between $g^{\gamma}_{1}$ and 
$F^{\gamma}_{1}$.

In the next section, we discuss the general framework of eight virtual
photon structure functions and positivity constraints. In section III, we
present our calculation of squark contributions to the photon structure
functions and the numerical analysis is carried out. In section IV, we examine various aspects of the real photon structure functions, like inequality
between $g_1^\gamma$ and $F_1^\gamma$ and the vanishing 1st
moment sum rule.
In section V, we discuss a possible signature for the squark in 
the structure function $W_{TT}^\tau$.
The final section is devoted to conclusion.

\newpage

\section{Photon structure functions and positivity constraints}
We consider the virtual photon-photon forward scattering amplitude for
$\gamma(q)+\gamma(p)\rightarrow \gamma(q)+\gamma(p)$  illustrated in Fig.1,
\be
T_{\mu\nu\rho\sigma}(p,q)=i\int d^4 x d^4 y d^4 z e^{iq\cdot x}e^{ip\cdot (y-z)}
\langle 0|T(J_\mu(x) J_\nu(0) J_\rho(y) J_\sigma(z))|0\rangle~,
\ee
where $J$ is the electromagnetic current,  $q$ and $p$ are the four-momenta of 
the probe and target photon, respectively. The $s$-channel helicity amplitudes are related to its absorptive part as follows:
\be
W(ab\vert a'b')=\epsilon^*_\mu(a)\epsilon^*_\rho(b)
W^{\mu\nu\rho\sigma}\epsilon_\nu(a')\epsilon_\sigma(b')~,
\ee
where
\be
W_{\mu\nu\rho\sigma}(p,q)=\frac{1}{\pi}{\rm Im}T_{\mu\nu\rho\sigma}(p,q)~,
\ee
and $\epsilon_\mu (a)$ represents the photon polarization vector with helicity 
$a$, and $a =0, \pm1$. Similarly for the other polarization vectors and we have  $a', b, b'=0, \pm1$. Due to the angular momentum conservation, parity conservation and time reversal invariance \cite{BLS}, we have in total eight independent $s$-channel helicity amplitudes, which we may take as
\bea
&&W(1,1\vert 1,1),\ \
W(1,-1\vert 1,-1),\ \  W(1,0\vert 1,0),\ \  W(0,1\vert 0,1),\ \  W(0,0\vert
0,0),
\nonumber  \\
&&W(1,1\vert -1,-1),\ \  W(1,1\vert 0,0),\ \  W(1,0\vert 0,-1).
\eea
The first five amplitudes are helicity-nonflip  and the last three are
helicity-flip. It is noted that the $s$-channel helicity-nonflip amplitudes
are semi-positive, but not the helicity-flip ones.

\begin{figure}[hbt]
\begin{center}
\includegraphics[scale=0.4]{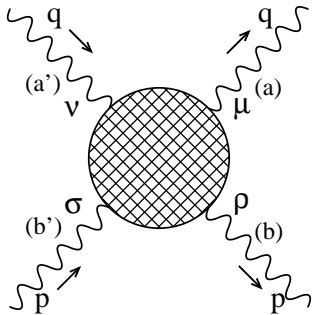}
\vspace{-0.5cm}
\caption{\label{virtual-forward}Virtual photon-photon forward scattering with momenta $q(p)$ and helicities $a(b)$ and $a'(b')$ } 
\end{center}
\end{figure}

In our previous works~\cite{SSU1,SSU2}, 
we have applied the Cauchy-Schwarz inequality~\cite{JS,JST}  to the above photon helicity amplitudes and have
derived a positivity bound:
\be
\Bigl|W(a,b\vert a',b')  \Bigr|\leq \sqrt{W(a,b\vert a,b)
W(a',b'\vert a',b')}~.
\ee
Writing down explicitly, we obtain
the following three positivity constraints:
\bea
\Bigl|W(1,1\vert -1,-1)  \Bigr|&\leq& W(1,1\vert 1,1)~,  \label{CS1}\\
\Bigl|W(1,1\vert 0,0)  \Bigr|&\leq& \sqrt{W(1,1\vert 1,1)
W(0,0\vert 0,0)}~,\label{CS2}\\
\Bigl|W(1,0\vert 0,-1)  \Bigr|&\leq& \sqrt{W(1,0\vert 1,0)W(0,1\vert 0,1)}~.
\label{CS3}
\eea

The photon-photon scattering phenomenology is often discussed in terms of
the photon structure functions instead of the $s$-channel helicity amplitudes.
Budnev, Chernyak and Ginzburg [BCG]
\cite{BCG}  introduced the following eight independent structure functions, in
terms of
which  the absorptive part of virtual photon-photon forward scattering,
$W^{\mu\nu\rho\sigma}$, is written as (See Appendix A),
\begin{eqnarray}
W_{\mu\nu\rho\sigma}&=&(T_{TT})_{\mu\nu\rho\sigma} W_{TT}
+(T^{a}_{TT})_{\mu\nu\rho\sigma} W^{a}_{TT}
+(T^{\tau}_{TT})_{\mu\nu\rho\sigma} W^{\tau}_{TT}+
(T_{LT})_{\mu\nu\rho\sigma} W_{LT}\nonumber\\
&&+(T_{TL})_{\mu\nu\rho\sigma} W_{TL}+(T_{LL})_{\mu\nu\rho\sigma} W_{LL}
-(T^{\tau}_{TL})_{\mu\nu\rho\sigma} W^{\tau}_{TL}
-(T^{\tau a}_{TL})_{\mu\nu\rho\sigma} W^{\tau a}_{TL}~,
\label{had1}
\end{eqnarray}
where $T_i$'s are the projection operators given in Appendix A.

The virtual photon structure functions $W_i$'s are functions of
three invariants, i.e., $p\cdot q$, $q^2(=-Q^2)$ and $p^2(=-P^2)$, and have no
kinematical singularities.
The subscript \lq\lq$ T$\rq\rq\  and \lq\lq$ L$\rq\rq\  refer
to the transverse and longitudinal photon, respectively.
The structure functions with the superscript \lq\lq $\tau$\rq\rq\
correspond to transitions with spin-flip for each of the photons with
total helicity conservation, while those with the superscript  \lq\lq 
$a$\rq\rq\
correspond to the $\mu\nu$ antisymmetric part of $W_{\mu\nu\rho\sigma}$
and are measured, for example,  through  the two-photon processes in polarized
$e^+e^-$ collision experiments.
These eight structure functions are related to the $s$-channel helicity amplitudes as follows~\cite{BCG}:
\bea
W_{TT}&=&\frac{1}{2}\left[W(1,1\vert 1,1)+W(1,-1\vert 1,-1)   \right]~,
\qquad W_{LT}= W(0,1\vert 0,1)~,  \nonumber \\
W_{TL}&=&W(1,0\vert 1,0)~, \qquad W_{LL}=W(0,0\vert 0,0)~,    \nonumber \\
W^a_{TT}&=&\frac{1}{2}\left[W(1,1\vert 1,1)-W(1,-1\vert 1,-1)   \right]~,
\qquad W^{\tau}_{TT}= W(1,1\vert -1,-1)~,  \nonumber \\
W^{\tau}_{TL}&=& \frac{1}{2}\left[W(1,1\vert 0,0)-W(1,0\vert 0, -1)   \right]~,
\nonumber
\\ W^{\tau a}_{TL}&=&\frac{1}{2}\left[W(1,1\vert 0,0)+W(1,0\vert 0, -1)
\right]~. \label{HelicityVSBCG}
\eea
Since the helicity-nonflip amplitudes are non-negative, the first four
structure functions are  positive semi-definite and the last four are not. 
Due to the
fact that the
absorptive part $W_{\mu\nu\rho\sigma}(p,q)$  is symmetric under the simultaneous
interchange of $\{q,\mu,\nu\}\leftrightarrow \{p,\rho,\sigma\}$, all the
virtual photon structure functions, except $W_{LT}$ and $W_{TL}$, are symmetric
under interchange of $p \leftrightarrow q$, while $W_{LT}(p\cdot q,q^2,p^2)
=W_{TL}(p\cdot q,p^2,q^2)$.
In terms of these structure functions, the positivity constraints
(\ref{CS1})-(\ref{CS3}) are rewritten as
\bea
\Bigl|W_{ TT}^\tau \Bigr|&\leq&
\left(W_{ TT}+W_{ TT}^a\right)~,  \label{BCG1}\\
\Bigl|W_{ TL}^\tau +W_{ TL}^{\tau a}  \Bigr|&\leq&
\sqrt{(W_{ TT}+W_{ TT}^a)W_{ LL}}~,\label{BCG2}\\
\Bigl| W_{ TL}^\tau -W_{ TL}^{\tau a} \Bigr|&\leq&
\sqrt{W_{ TL}W_{ LT}}\label{BCG3}~.
\eea
In fact,  the following bounds,
\be
         \Bigl|W_{ TT}^\tau \Bigr|\leq 2 W_{ TT}~, \qquad
2\Bigl( W_{ TL}^\tau  \Bigr)^2 \leq 2W_{LL} W_{ TT}+W_{ TL}W_{ LT}~,
\label{oldCostraint}
\ee
were derived, some time ago, from the positiveness of the $\gamma\gamma$
cross-section for arbitrary  photon polarization \cite{BGM}. Note that the
 constraints (\ref{BCG1})-(\ref{BCG3}) which we have obtained
are more stringent than the above ones (\ref{oldCostraint}).

\section{Calculation of squark contribution and the results}
The structure functions are evaluated by multiplying the 
relevant projection operator to the structure tensor
$W_{\mu\nu\rho\sigma}$ which is the imaginary part of the 
the forward photon-photon amplitude $T_{\mu\nu\rho\sigma}$:
\bea
W_i=P_i^{\mu\nu\rho\sigma}\frac{1}{\pi}{\rm Im}T_{\mu\nu\rho\sigma}
=\int dPS^{(2)}P_i^{\mu\nu\rho\sigma}{\cal M}^{*}_{\mu\rho}
{\cal M}_{\nu\sigma}~,
\eea
where $P_i$'s are the normalized projection operators defined
in Appendix A.
In our calculation, we evaluated the structure functions 
by two methods; (i) Computing the discontinuity of the forward
photon-photon amplitude (see Fig.2) multiplied by projection operators, and
(ii) Integrating the squared amplitudes 
$\cal{{M}^{*}_{\mu\rho}}{\cal M}_{\nu\sigma}$ for the squark $\tilde{q}$
and anti-squark $\bar{\tilde{q}}$ production 
$\gamma+\gamma\rightarrow\tilde{q}+\bar{\tilde{q}}$,
multiplied by projection
operators over the two-body phase space $dPS^{(2)}$. 
Both calculations coincide for the eight structure functions.
\begin{figure}[hbt]
\begin{center}
\includegraphics[scale=0.3]{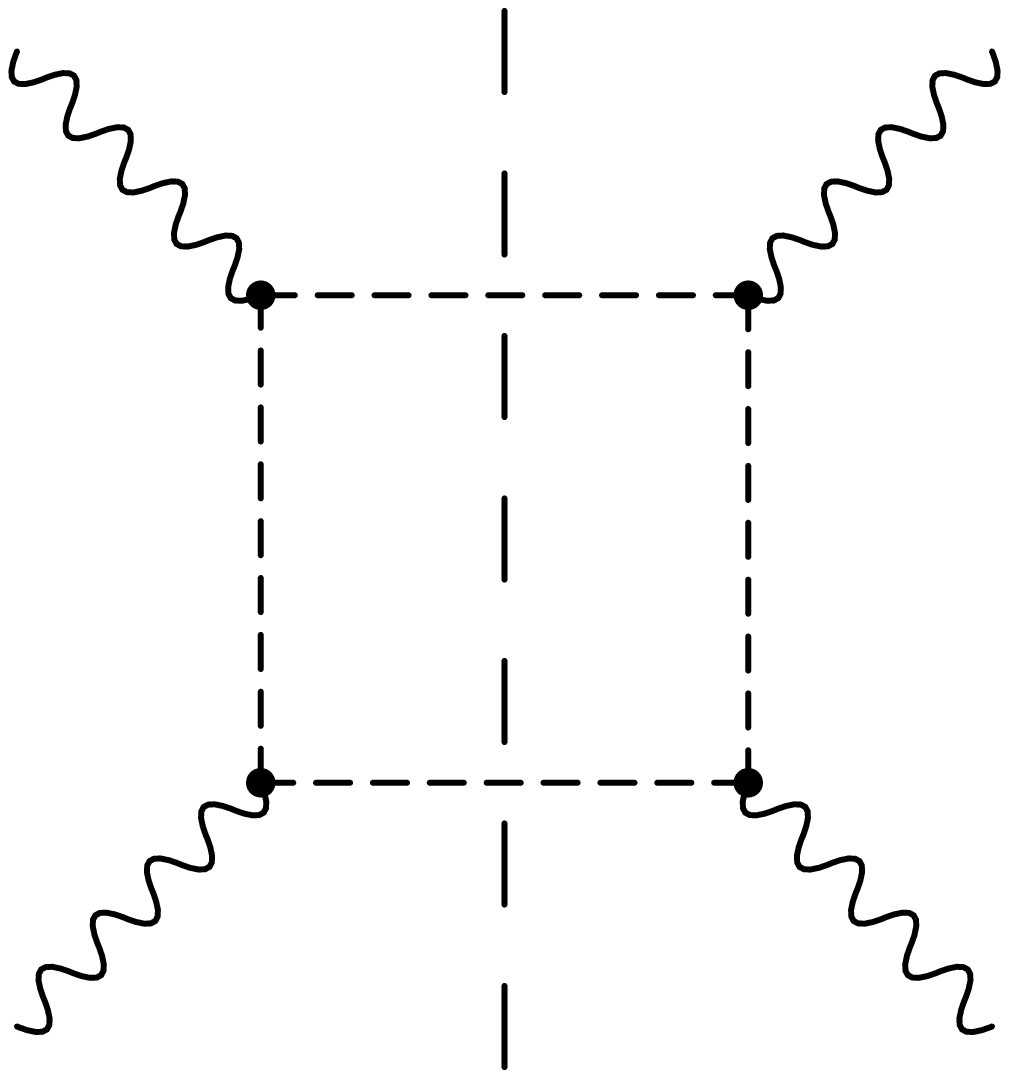}
\hspace{1cm}
\includegraphics[scale=0.3]{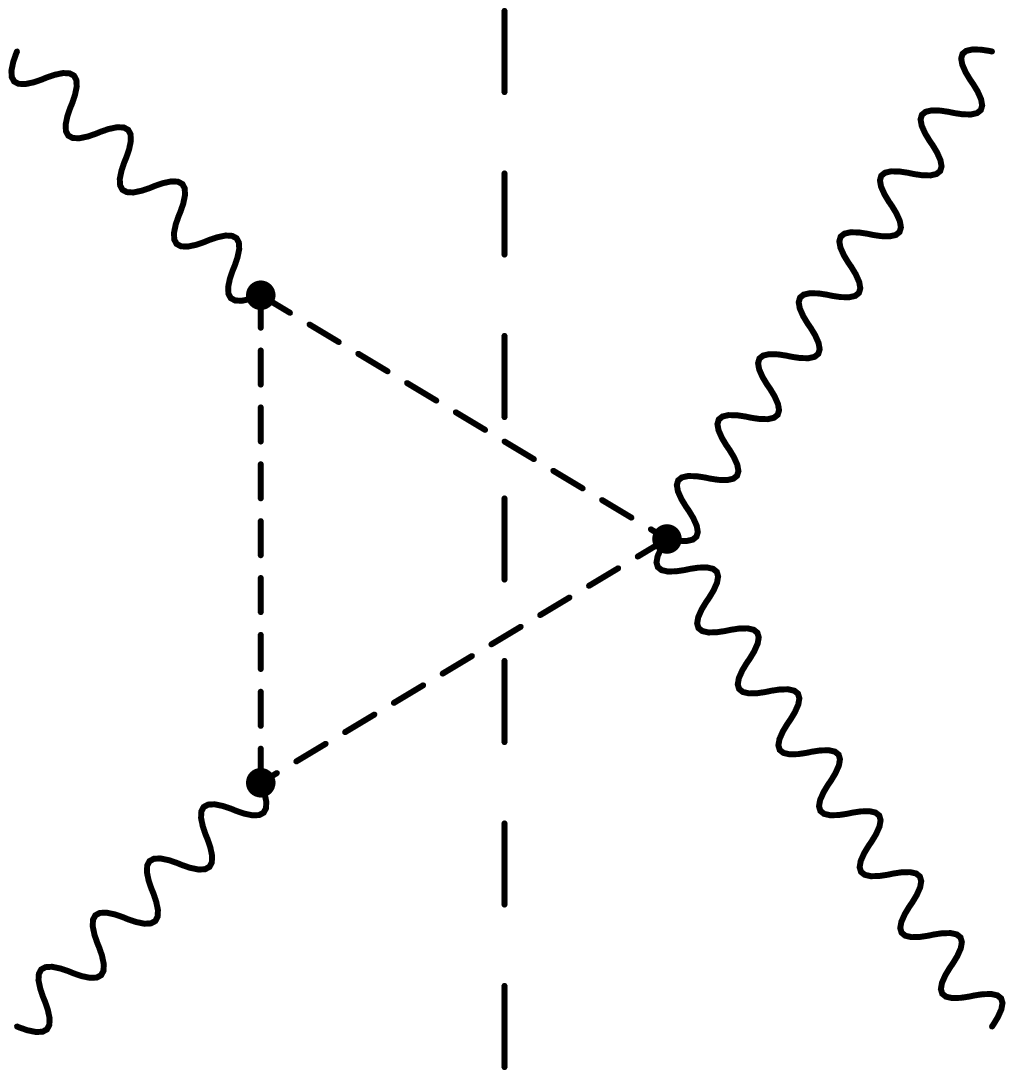}
\hspace{1cm}
\includegraphics[scale=0.4]{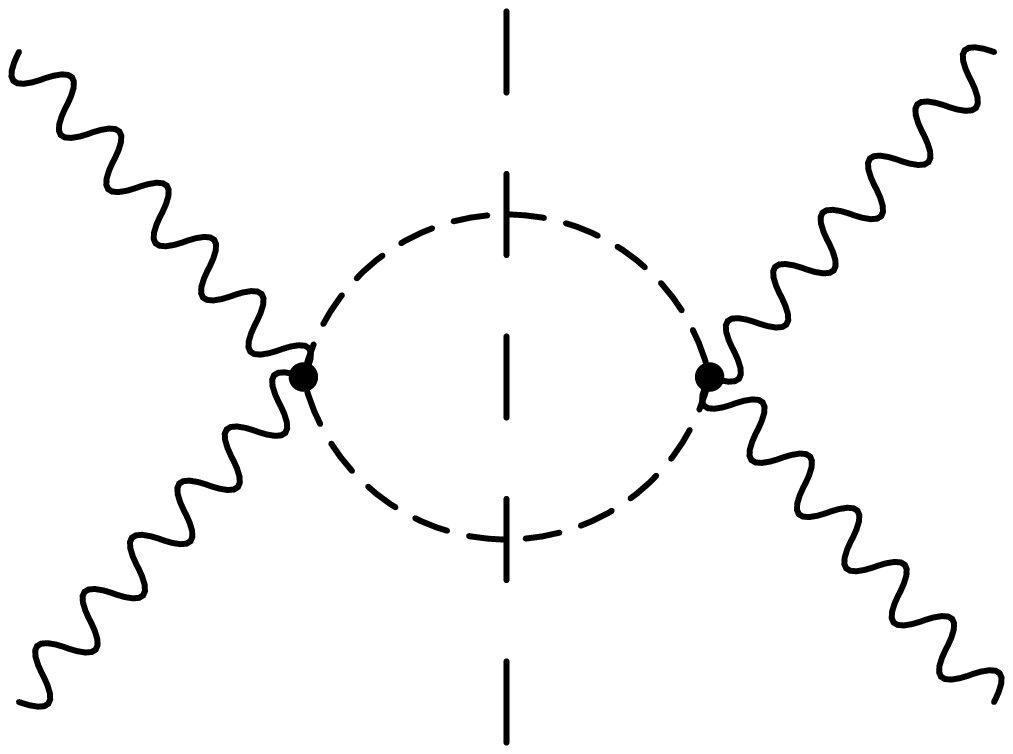}
\caption{\label{squark-box} 
Squark-loop diagrams (box, triangle, bubble) contribution 
to photon structure functions} 
\end{center}
\end{figure}

We have summarized our results for the eight virtual structure functions
in the Appendix B. Here we present the expressions converted to
the structure functions usually used for the nucleon target in the
following.

We note that the virtual photon structure functions $F_1^\gamma$, 
$F_2^\gamma$, $F_L^\gamma$, $g_1^\gamma$ and $g_2^\gamma$
are related to the ones  introduced by BCG  in \cite{BCG} as 
follows:
\bea
F_1^\gamma(x,Q^2,P^2)&=& W_{TT}-\frac{1}{2}W_{TL}, \nonumber \\
F_2^\gamma(x,Q^2,P^2)&=&\frac{x}{{\widetilde\beta}^2} \left[
   W_{TT}+W_{LT}-\frac{1}{2}W_{LL} -\frac{1}{2}W_{TL}  \right], \nonumber \\
F_L^\gamma(x,Q^2,P^2)&=&F_2^\gamma-xF_1^\gamma,   \\
g_1^\gamma(x,Q^2,P^2)&=&\frac{1}{{\widetilde\beta}^2}\left[W^a_{TT}-
\sqrt{1-\tbeta^2}W^{\tau a}_{TL}
     \right],  \nonumber\\
g_2^\gamma(x,Q^2,P^2)&=&-\frac{1}{{\widetilde\beta}^2}\left[W^a_{TT}-
\frac{1}{\sqrt{1-\tbeta^2}}W^{\tau a}_{TL}
     \right],  \nonumber
\eea
where independent variables are $x=Q^2/2p\cdot q$ (Bjorken variable),
$P^2$, $Q^2$ and $m^2$. Here we have introduced the variable $\tbeta$
given as
\bea
&&\widetilde{\beta}=\sqrt{1-\frac{P^2Q^2}{(p\cdot q)^2}}=
\sqrt{1-\frac{4x^2P^2}{Q^2}}~.
\eea
In order to write down above structure functions, we also introduce 
the following variables:
\bea
&&\beta=\sqrt{1-\frac{4m^2}{(p+q)^2}}=\sqrt{1+\frac{4m^2 x}{xP^2+
(x-1)Q^2}}~,\\
&&L=\ln\frac{1+\beta\tbeta}{1-\beta\tbeta}~.
\eea
One of the characteristics of the squark diagrams,   
$W_{LT}$ and $W_{TL}$ do not receive any contribution from
the triangle and bubble diagrams which consist of seagull
graphs \cite{SS}.

For a flavor $q$ the structure functions turn out to be
\bea
&&F_1^\gamma=N_c\frac{\alpha}{\pi}e_q^4\nn\\
&&\times\left[-\frac{1}{\tbeta^3}L
\left\{-8\tbeta^2\frac{m^4}{Q^4}x^2+\frac{m^2}{Q^2}\left[
(1-\tbeta^2)^2+\frac{1}{2}(1-\tbeta^2)(12x^2-4x-3)
+4x(1-2x)\right]
\right.\right.\nn\\
&&\left.\left.\hspace{1cm}+\frac{1}{8}
\frac{P^2}{Q^2}\left((1-\tbeta^2)+4x^2-4x\right)
\left((1-\tbeta^2)+8x^2-2\right)\right\}
+\frac{\beta}{\tbeta^2}\left\{
\frac{P^2}{Q^2}x(1-3x)\right.
\right.\nn\\
&&\left.\left.\hspace{1cm}+\frac{4m^2}{Q^2}x(1-x)
-\frac{1}{4x}(1-\tbeta^2)\left[(1-x)(1-\beta^2)
+x\beta^2\frac{P^2}{Q^2}\right]-2x(1-x)+1\right\}\right]~,\\
&&F_2^\gamma=N_c\frac{\alpha}{\pi}e_q^4 x\nn\\
&&\times\left[
\frac{1}{\tbeta^5}L\left\{8\tbeta^2\frac{m^4}{Q^4}x^2
+\frac{m^2}{Q^2}
\left[\frac{1}{2}(1-\tbeta^2)(-12x^2-4x+3)+4x(3x-1)\right]
+\frac{1}{8}\frac{P^2}{Q^2}(1-\tbeta^2)^2
\right.\right.\nn\\
&&\left.\left.\hspace{1cm}
+\frac{1}{4}\frac{P^2}{Q^2}(1-\tbeta^2)(2x^2-2x+1)
-\frac{P^2}{Q^2}x\left(2x(2x^2-2x+1)+1\right)+2x(1-x)
\right\}\right.\nn\\
&&\left.\hspace{1cm}
+\frac{\beta}{2\tbeta^4(1-\beta^2\tbeta^2)}
\left\{2(\beta^2-1)\tbeta^2\frac{m^2}{Q^2}
\left[(1-\tbeta^2)-4x(1-x)\right]
+2(1-\beta^2)\left[\frac{P^2}{Q^2}(1-\tbeta^2)^2
\right.\right.\right.\nn\\
&&\left.\left.\left.\hspace{1cm}
+\frac{1}{4}\frac{P^2}{Q^2}(1-\tbeta^2)
(12x^2-28x-1)+\frac{P^2}{Q^2}x(1-x)(2x+1)(10x+1)
-8x(1-x)+1\right]\right.\right.\nn\\
&&\left.\left.\hspace{1cm}
+(1-\tbeta^2)\left[-\frac{3}{2}\frac{P^2}{Q^2}(1-\tbeta^2)
+\frac{2P^2}{Q^2}x(x+3)+(12x^2-12x+1)\right]\right\}\right]~,
\label{F2-virtual}
\eea
\bea
&&F_L^\gamma=N_c\frac{\alpha}{\pi}e_q^4\nn\\
&&\times\left[
-\frac{1}{2\tbeta^3}L(2x-1+\tbeta^2)
\left\{\frac{1}{4}(1-\tbeta^2)(4x-1)-\frac{4m^2}{Q^2}x^2
+\frac{1}{2}(1-\tbeta^2)\left[(1-x)(1-\beta^2)
+x\beta^2\frac{P^2}{Q^2}\right]
\right.\right.\nn\\
&&\left.\left.\hspace{1cm}-2x(1-x)\right\}
+\frac{2\beta x}{\tbeta^2(1-\beta^2\tbeta^2)}
\left\{-\frac{m^2}{Q^2}\left[(1-\tbeta^2)^2-4(1-\tbeta^2)
x(2x+1)+12x^2\right]
\right.\right.\nn\\
&&\left.\left.\hspace{1cm}-\frac{P^2}{Q^2}\left[
\frac{1}{4}(1-\tbeta^2)^2-(1-\tbeta^2)x(2x+1)+x^2+8x^3(1-x)\right]
\right\}\right]~,\\
&&g_1^\gamma=N_c\frac{\alpha}{\pi}e_q^4\nn\\
&&\times\left[
\frac{1}{\tbeta^5}L\left\{
\frac{2m^2x}{Q^2}\left[(1-\tbeta^2)^2-3(1-\tbeta^2)+2\right]
+\frac{P^2}{Q^2}x\left[\frac{1}{2}(1-\tbeta^2)^2
+2(1-\tbeta^2)(x^2-1)\right.\right.\right.\nn\\
&&\left.\left.\left.\hspace{1cm}-2x(4x-3)\right]\right\}
+\frac{\beta}{\tbeta^4}\left\{
\frac{P^2}{Q^2}x(1-\tbeta^2)+\frac{2P^2x}{Q^2}(2x^2-4x+1)
+(2x-1)\right\}
\right]~,\\
&&g_2^\gamma=N_c\frac{\alpha}{\pi}e_q^4\nn\\
&&\times\left[
\frac{1}{2\tbeta^2}L\left\{
\frac{P^2}{Q^2}2x(2x^2-4x+1)-\frac{4m^2x}{Q^2}+
(1-\tbeta^2)\left[(1-x)(1-\beta^2)+x\beta^2\frac{P^2}{Q^2}\right]
+(2x-1)\right\}\right.\nn\\
&&\left.\hspace{2cm}+\frac{\beta}{\tbeta^4}
\left\{\frac{P^2}{Q^2}x(4x-3)+(2-3x)\right\}\right]~,
\eea
where $\alpha=e^2/4\pi$ is the fine structure constant of QED,
$N_c$ is the number of colors, $N_c=3$ for supersymmetric QCD.
$e_q$ is the electric charge of squark of $q$-th flavor. 
In order to take into account all the flavor contributions we 
have to sum over flavors $\sum_q$.

We should note that the variables $L$, $\beta$ and $\tbeta$ 
are not independent of the variables $x$, $P^2$, $Q^2$ and $m^2$,
the expressions given here are not unique. Also we should note that 
these structure functions do not depend on the dimensionful variables
$Q^2$, $P^2$ and $m^2$, but they depend only on the ratios, $P^2/Q^2$ 
and $m^2/Q^2$.

\begin{figure}[hbt]
\begin{center}
\begin{tabular}{cc}
\includegraphics[scale=0.6]{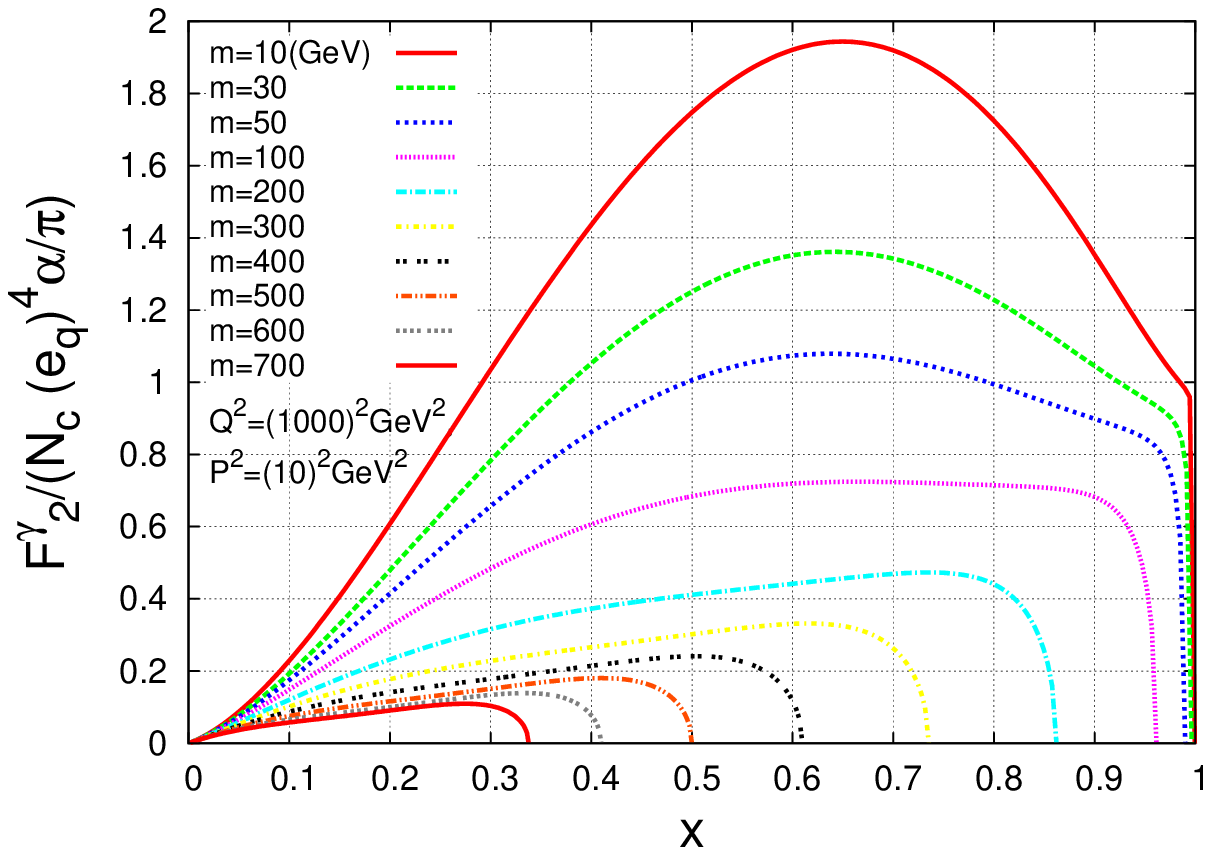}&
\includegraphics[scale=0.6]{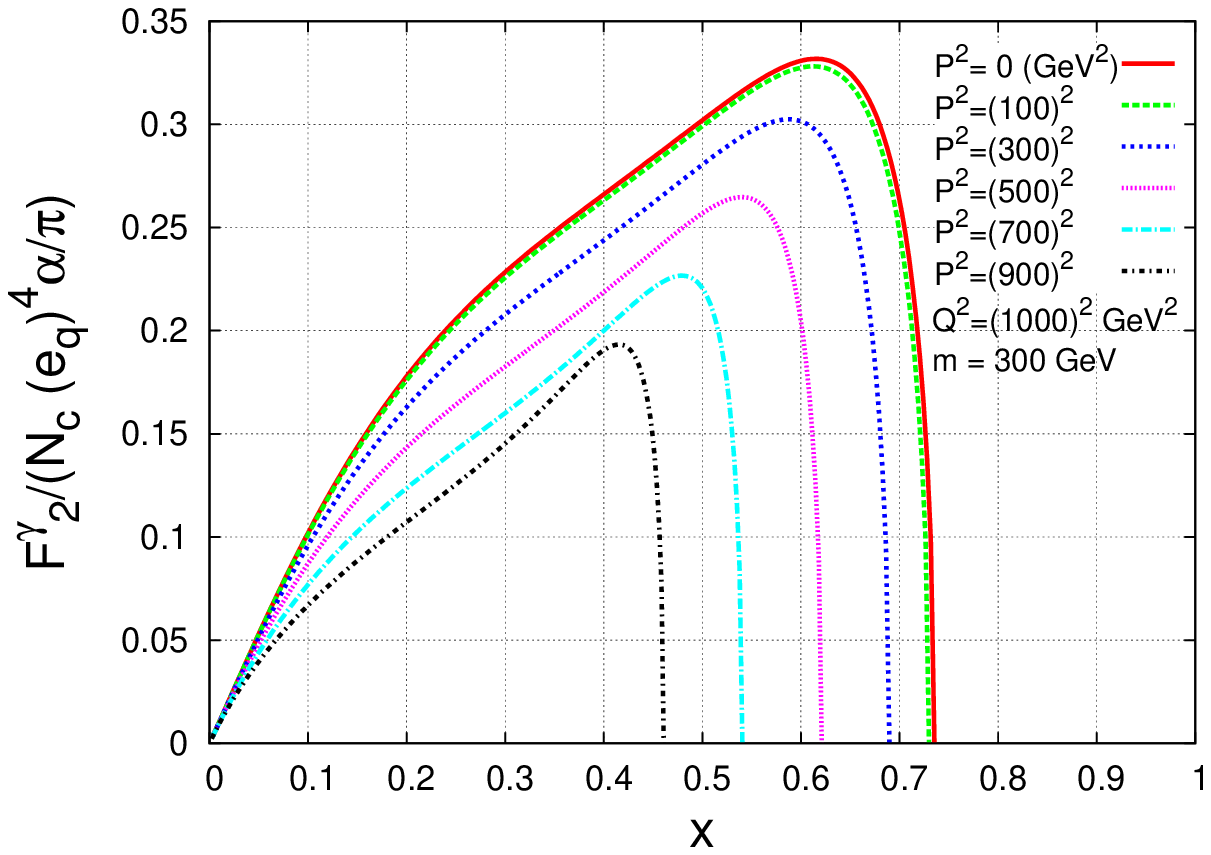}\\
\end{tabular}
\caption{\label{F2}$F_2^\gamma$ as a function of $x$ for various values
of squark mass, $m$, which are given in units of GeV, for fixed
$P^2=(10)^2$ GeV$^2$ and $Q^2=(1000)^2$ GeV$^2$ (Left), and for various
$P^2$ in units of GeV$^2$. with $m=300$GeV (Right).
}
\end{center}
\end{figure}
\vspace{-0.3cm}
\begin{figure}[hbt]
\begin{center}
\begin{tabular}{cc}
\includegraphics[scale=0.6]{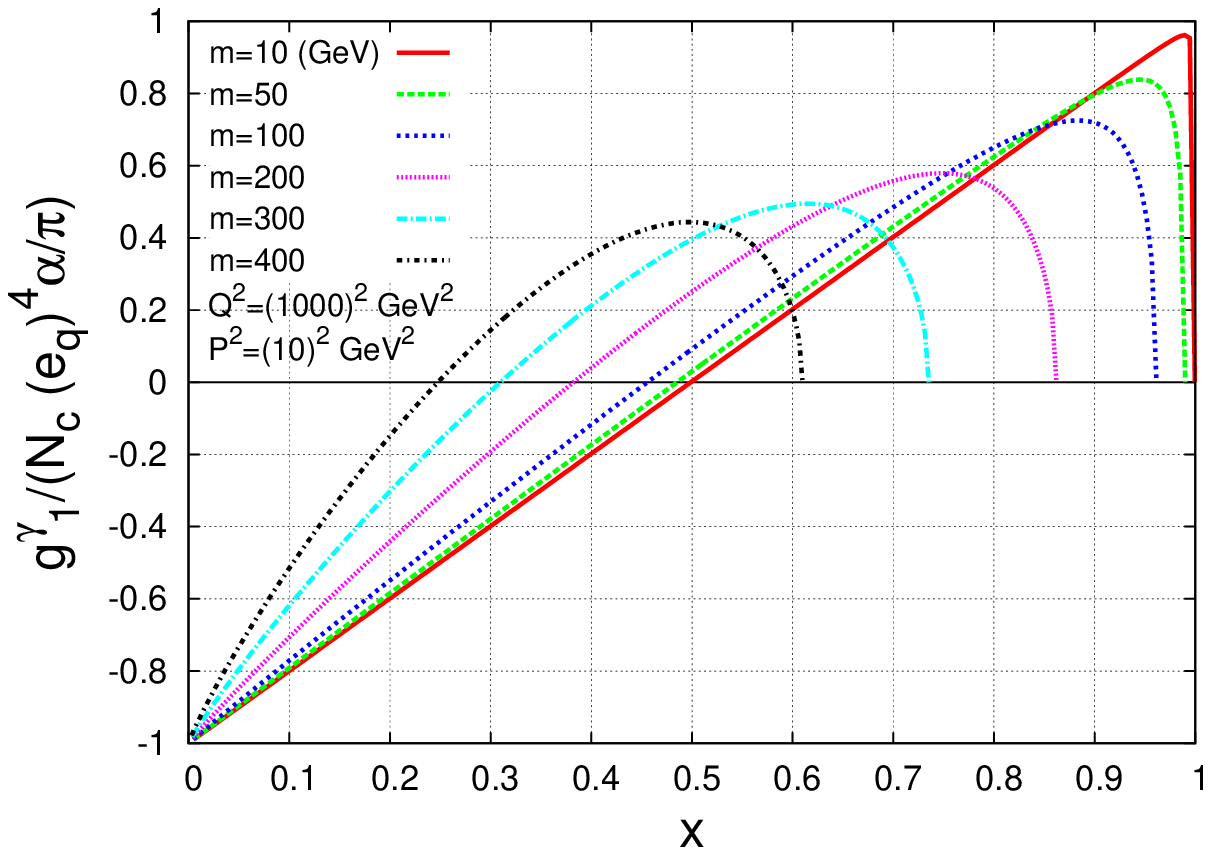}&
\includegraphics[scale=0.6]{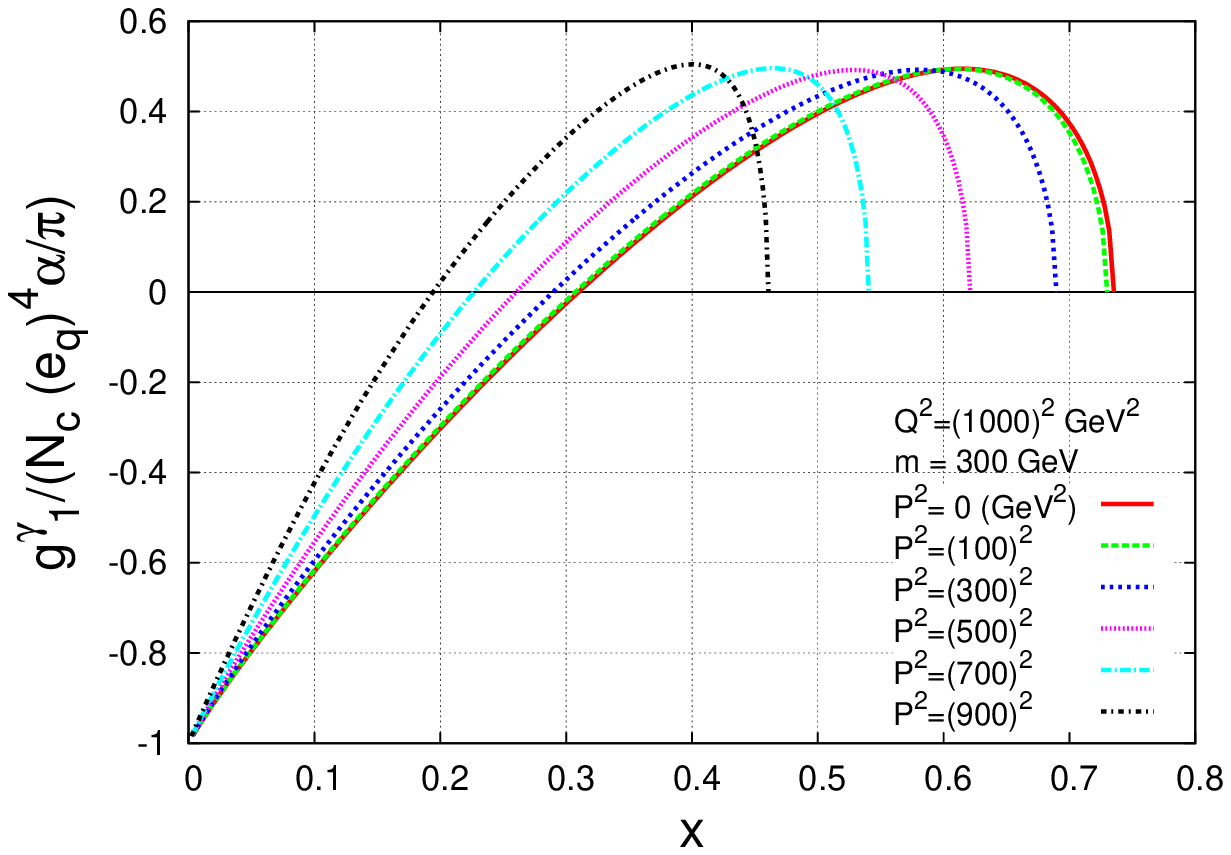}\\
\end{tabular}
\caption{\label{g1}$g_1^\gamma$ for various squark mass $m$ 
in units of GeV
(Left), and for various $P^2$ in units of GeV$^2$ with fixed
$Q^2=(1000)^2$ GeV$^2$ and $m=300$ GeV (Right).}
\end{center}
\end{figure}

\begin{figure}[hbt]
\begin{center}
\includegraphics[scale=0.6]{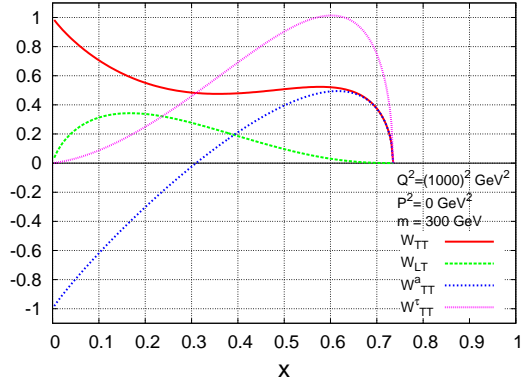}
\caption{\label{Wset1R}Four real photon structure functions:$W_{TT}$, $W_{LT}$, $W^a_{TT}$, $W^\tau_{TT}$}
\end{center}
\end{figure}
\begin{figure}[hbt]
\begin{center}
\begin{tabular}{cc}
\includegraphics[scale=0.6]{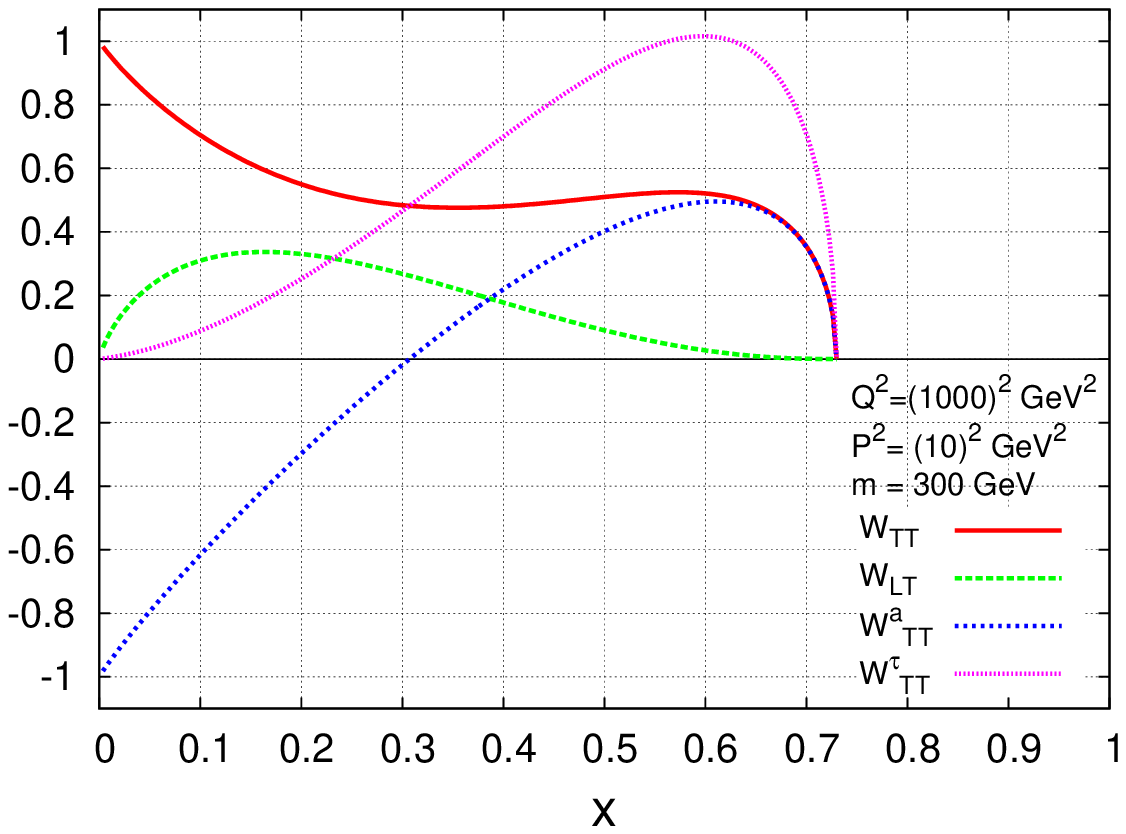}&
\includegraphics[scale=0.6]{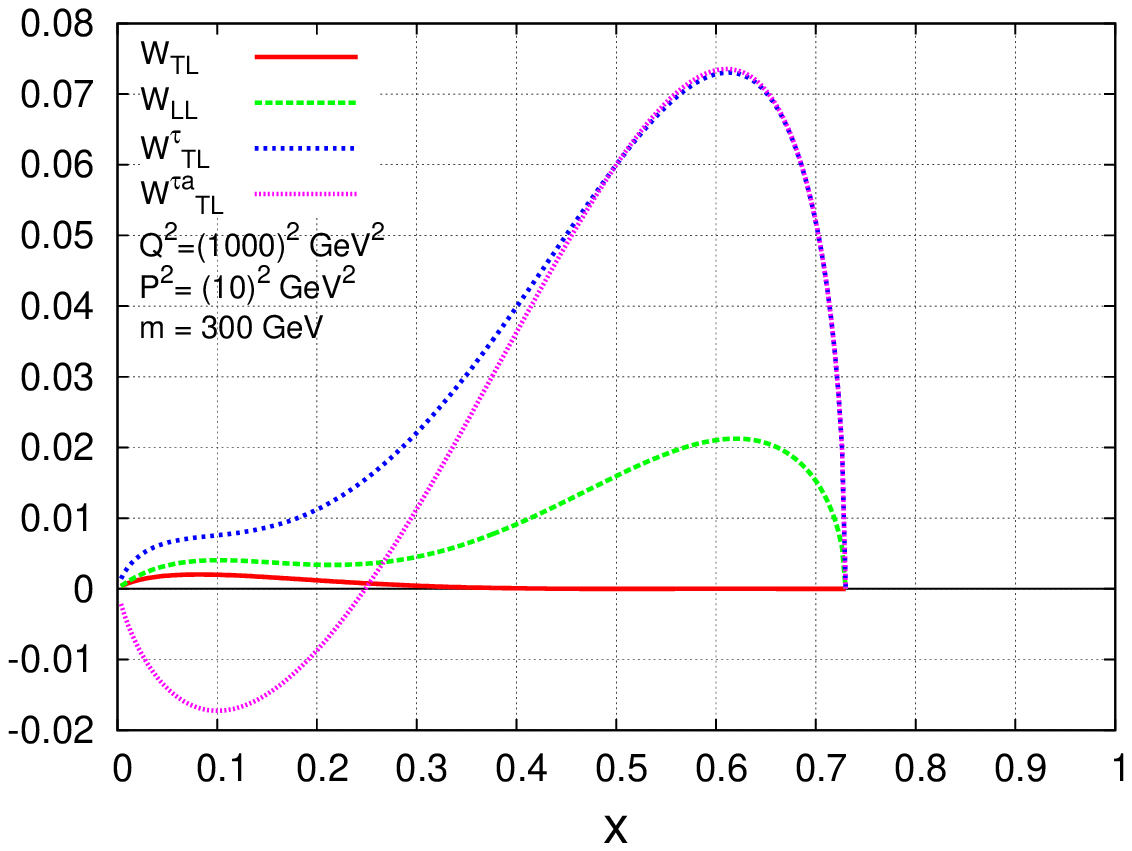}\\
\end{tabular}
\caption{\label{WV}Eight virtual photon structure functions:$W_{TT}$, $W_{LT}$, $W^a_{TT}$, $W^\tau_{TT}$(Left), $W_{TL}$, $W_{LL}$, $W^\tau_{TL}$, $W^{\tau a}_{TL}$(Right). The values of $P^2$ and $Q^2$ are in units of GeV$^2$.}
\end{center}
\end{figure}
We plot in Figs. \ref{F2}, \ref{g1}, \ref{Wset1R} and \ref{WV},
squark contributions to the photon structure functions as
functions of $x$. The vertical axes are in units of 
$N_c\frac{\alpha}{\pi}e_q^4$, where $N_c$ is the number of 
colors, $N_c=3$ for supersymmetric QCD. $e_q$ is the electric
charge of the squark which is the super partner of the
quark of the $q$-th flavor.

In these plots we have chosen $Q^2=(1000)^2$GeV$^2$, and
$P^2=(10)^2$GeV$^2$. The allowed $x$ region is
$0\leq x\leq x_{\rm max}$ with
\bea
x_{\rm max}=\frac{1}{1+\frac{P^2}{Q^2}+\frac{4m^2}{Q^2}}~.
\eea
The photon structure functions can be classified
into two groups: (i) $W_{TT}$, $W_{LT}$, $W_{TT}^a$, $W_{TT}^\tau$
and (ii) $W_{TL}$, $W_{LL}$, $W_{TL}^\tau$, $W_{TL}^{\tau a}$.
The first group also exists for the real photon target, while the
second group does not exist for the real photon case and are small
in magnitude compared to the first group. The graphs show that
all the structure functions tend to vanish as 
$x\rightarrow x_{\rm max}$ which is the kinematical constraint.

\subsection{Positivity and Equality}
The positivity constraints (\ref{BCG1}) and (\ref{BCG3}) derived from the 
general Cauchy-Schwarz inequalities in fact lead to the following equalities
for the squark contributions:
\bea
&&W_{TT}^\tau=W_{TT}+W_{TT}^a\label{equality1}~,\\
&&|W_{TL}^\tau-W_{TL}^{\tau a}|=\sqrt{W_{TL}W_{LT}}~,\label{equality2}
\eea
while we have an inequality
\bea
\Bigl|W_{ TL}^\tau +W_{ TL}^{\tau a}  \Bigr|\leq
\sqrt{(W_{ TT}+W_{ TT}^a)W_{ LL}}~.\label{WTL-ineq}
\eea
The first equality (\ref{equality1}) can be rewritten in terms
of the helicity amplitudes as
\bea
W(1,1|-1,-1)=W(1,1|1,1)~,\label{equality-flip}
\eea
which holds both for the real ($P^2=0$) and virtual
($P^2\neq 0$) photon target. We can also read off this relation
from the Figs. \ref{Wset1R} and \ref{WV}.
In the limit $x\rightarrow 0$, for example, $W_{TT}\rightarrow 1$, while
$W_{TT}^a\rightarrow -1$ and hence $W_{TT}^\tau\rightarrow 0$.
Note that because of Eq.(\ref{equality-flip}), 
$W(1,1|-1,-1)$ or $W_{TT}^\tau$ is positive definite, and
the left-hand side of (\ref{equality1}) is without an absolute
value symbol.

The second equality (\ref{equality2})
only exists for the virtual photon case.
One can also see that this relation holds from the Fig. \ref{WV}, 
where
$W_{TL}^\tau$ almost overlaps with $W_{TL}^{\tau a}$ at larger $x$
for which the product $W_{TL}W_{LT}$ looks very small, while
in the smaller $x$ region the difference $W_{TL}^\tau-W_{TL}^{\tau a}$
becomes sizable and the product $W_{TL}W_{LT}$ shows non-vanishing
values.

\begin{figure}[hbt]
\vspace*{-0.3cm}
\begin{center}
\includegraphics[scale=0.6]{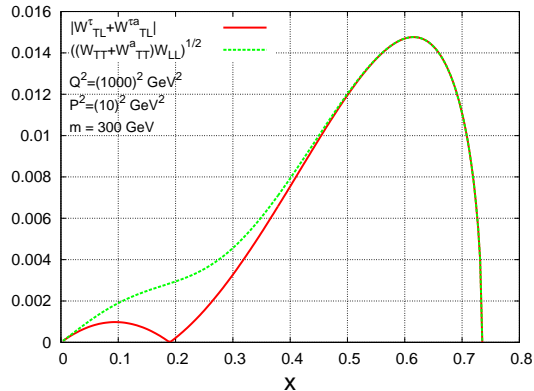}
\caption{\label{W-ineq}Inequality (\ref{WTL-ineq}) for $Q^2=(1000)^2$
GeV$^2$, $P^2=(10)^2$ GeV$^2$ and $m=300$ GeV.
}
\end{center}
\end{figure}

The inequality (\ref{WTL-ineq}) is illustrated in Fig.\ref{W-ineq}.
\newpage
\section{Real Photon Case}
We now consider the real photon case of the above structure functions
by taking the limit: $P^2\rightarrow 0$, or $\tbeta\rightarrow 1$.
Then the number of the independent structure functions reduces to
four for the real photon target. They are $W_{TT}$, $W_{LT}$, $W_{TT}^\tau$
and $W_{TT}^a$ given as follows: 
\bea
&&W_{TT}=N_c\frac{\alpha}{\pi}e_q^4
\left[L\ \tau x
\left\{\frac{1}{2}\tau x+(2x-1)\right\}
+\beta\left\{\tau x(1-x)+2x^2-2x+1\right\}\right]~,
\nn\\ \\
&&W_{LT}=N_c\frac{\alpha}{\pi}e_q^4 
\left[L\left\{\tau x^2+2x(1-x)\right\}
-6\beta x(1-x)\right]~,
\\
&&W_{TT}^a=N_c\frac{\alpha}{\pi}e_q^4
\left[L\ \tau x+\beta(2x-1)\right]~,
\\
&&W_{TT}^\tau=N_c\frac{\alpha}{\pi}e_q^4
\left[L\left\{2\tau\left(1+\frac{1}{4}\tau\right)x^2
\right\}+\beta\left\{2x^2+\tau x(1-x)\right\}\right]~,
\eea
where the logarithmic term $L$, the mass-ratio parameter $\tau$ and 
the velocity variable $\beta$ are defined for the real photon case as
\bea
L=\ln\frac{1+\beta}{1-\beta},\quad \tau=\frac{4m^2}{Q^2}, \quad
\beta=\sqrt{1-\frac{\tau x}{(1-x)}},\quad
\eea
which are different from $L$ and $\beta$ for the virtual photon case.
Note that from the above equation the following relation holds 
\bea
(1-x)(1-\beta^2)=\tau x~.
\eea
In terms of these four structure functions we can derive the
usual structure functions, $F_1^\gamma$, $F_2^\gamma$, $F_L^\gamma$
and $g_1^\gamma$ as follows:
\bea
&&F_1^\gamma=N_c\frac{\alpha}{\pi}e_q^4
\left[L\left\{
\frac{1}{2}\tau^2 x^2+\tau x(2x-1)\right\}
+\beta\left\{\tau x(1-x)+2x^2-2x+1\right\}
\right]~,\nn\\
\\
&&F_2^\gamma=N_c\frac{\alpha}{\pi}e_q^4\ x
\left[L
\left\{
\frac{1}{2}\tau^2x^2+\tau x(3x-1)+2x(1-x)
\right\}\right.\nn\\
&&\left.\hspace{6cm}+\beta\left\{\tau x(1-x)+8x^2-8x+1
\right\}\right]~,\label{F2real}\\
&&F_L^\gamma=N_c\frac{\alpha}{\pi}e_q^4\ x
\left[L\left\{\tau x^2
+2x(1-x)\right\}+6\beta (x^2-x)\right]~,\label{FLreal}
\\
&&g_1^\gamma=N_c\frac{\alpha}{\pi}e_q^4
\left[L\tau x
+\beta(2x-1)\right]~.
\eea
Note that we have the following relation:
\bea
F_L^\gamma=F_2^\gamma-xF_1^\gamma~.
\eea
Our result for $F_2^\gamma$ for the real photon
target (\ref{F2real}) is consistent with those in Refs.\cite{Reya,
SS,DGR,RW} and $F_L^\gamma$ in Ref.\cite{SS}
coincides with our result (\ref{FLreal}).
Note that our expression for $F_2^\gamma(x,Q^2,P^2)$ with 
$P^2\neq 0$ (\ref{F2-virtual}) is slightly different
from that given in Ref. \cite{RW}.
\subsection{Relation to the Splitting Functions}
It is well known that the collinear singularities
in the process of particle emission determine
the parton splitting functions and are related to the
$F_2^\gamma$ function. Namely the quark parton
distribution function inside the photon
reads in the leading logarithmic order
\bea
q^\gamma(x,Q^2)\sim P_{q\gamma}(x)\ln{Q^2/m^2}~,
\eea
where $P_{q\gamma}$ denotes the photon-quark splitting
function.  Then the structure function becomes 
\bea
F_{2,q}^\gamma\sim N_c \frac{\alpha}{\pi}\sum_q e_q^4
\ q^\gamma(x,Q^2)~.
\eea
Similarly for the squark contribution we have
\bea
&&F_{2,s}^\gamma\sim N_c \frac{\alpha}{\pi}\sum_s e_s^4
\ s^\gamma(x,Q^2)~,\\
&&s^\gamma(x,Q^2)\sim P_{s\gamma}(x)\ln{Q^2/m^2}~,
\eea
where we note that the splitting functions are given by
\cite{Kounnas-Ross,Jones-Smith}:
\bea
P_{q\gamma}(x)=x^2+(1-x)^2,\quad
P_{s\gamma}(x)=1-\left\{x^2+(1-x)^2\right\}=2x(1-x)~,
\eea
for which the following relation holds
\bea
P_{q\gamma}(x)+P_{s\gamma}(x)=1~.
\eea
\subsection{Mass singularities of the structure functions}
Let us consider the massless limit of the real photon
structure functions. Ignoring the power correction of
$m^2/Q^2$, the photon structure functions become
\bea
&& F_1^\gamma=W_{TT}\sim N_c\frac{\alpha}{\pi}e_q^4\left\{
2x^2-2x+1\right\}~,\\
&&F_2^\gamma=x\left[W_{TT}+W_{LT}\right]
\sim N_c\frac{\alpha}{\pi}e_q^4 x\left\{
2x(1-x)\ln\left(\frac{Q^2}{m^2}\frac{1-x}{x}\right)
+8x^2-8x+1\right\},\\
&&F_L^\gamma=xW_{LT}\sim N_c\frac{\alpha}{\pi}e_q^4 x\left\{
2x(1-x)\ln\left(\frac{Q^2}{m^2}\frac{1-x}{x}\right)
+6x(x-1)\right\}~,\\
&&g_1^\gamma=W_{TT}^a\sim N_c\frac{\alpha}{\pi}e_q^4\left\{
2x-1\right\}~.
\eea
In contrast to the spin $1/2$ quark, mass singularities
originate from $W_{LT}$, while $W_{TT}$
and $W_{TT}^a$ have no such singularities. Note that
for the spin $1/2$ quark case, such mass singularities
arise in $W_{TT}$ and $W_{TT}^a$.
This can be interpreted as the spinless nature of the
squark constituent.
In terms of the basis of $F_{1,2,L}^\gamma$, the mass singularity
appears in $F_2^\gamma$ and $F_L^\gamma$ for the squark case, 
in contrast to the
quark parton case, where mass singularities appear in 
$F_1^\gamma$ and $F_2^\gamma$.
Because of the logarithmic term due to mass singularities
of $F_L^\gamma$, the squark contribution to $F_L^\gamma$ 
is sizable compared to that for $F_2^\gamma$. 

\subsection{Inequality $|g_1^\gamma|\leq F_1^\gamma$}
\begin{figure}[hbt]
\begin{center}
\includegraphics[scale=0.7]{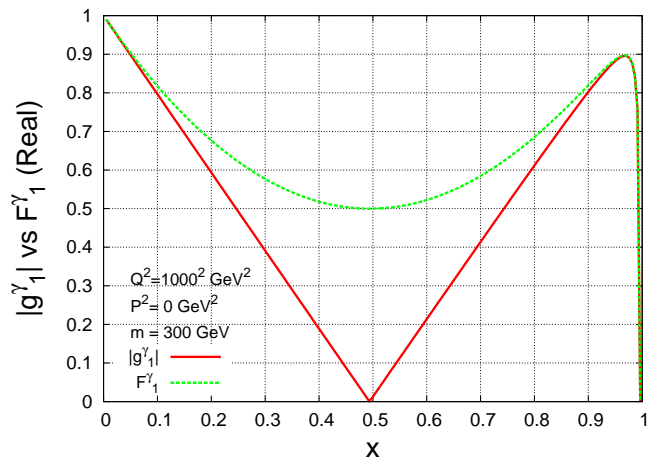}
\caption{\label{g1F1}
The inequality:\ $|g_1^\gamma|\leq F_1^\gamma$ for the real
photon target in the case of $Q^2=(1000)^2$GeV$^2$ and $m=300$GeV.}
\end{center}
\end{figure}

For the real photon 
\bea
&&g_1^\gamma=W_{TT}^a=\frac{1}{2}\left[
W(1,1|1,1)-W(1,-1|1,-1)\right]~,\\
&&F_1^\gamma=W_{TT}=\frac{1}{2}\left[
W(1,1|1,1)+W(1,-1|1,-1)\right]~.
\eea
Since the helicity non-flip amplitudes $W(1,1|1,1)$
and $W(1,-1|1,-1)$ are semi-positive definite, we are
led to the inequality:
\bea
|g_1^\gamma|\leq F_1^\gamma~,
\eea
which holds both for the squark and quark contribution.
This can be shown in the Fig.\ref{g1F1}. 

We have also numerically
studied the above inequality for the virtual photon case
and it has turned out that the inequality is not satisfied at
small $x$ region for the case where $P^2$ is much bigger than
$m^2$.

\subsection{$g_1^\gamma$ sum rule}
For the squark contribution
the 1st moment of the $g_1^\gamma$ structure functions turns
out to be
\bea
\int_0^{x_{\rm max}}g_1^\gamma(x,Q^2)dx
=N_c\frac{\alpha}{\pi}e_q^4
\left[\int_0^{x_{\rm max}}
\tau xLdx +\int_0^{x_{\rm max}}\beta(2x-1)dx\right]~,
\eea
where 
\bea
\tau=\frac{4m^2}{Q^2},\quad x_{\rm max}=\frac{1}{1+\tau}~.
\eea
Now by repeated use of integration by parts, where we get
vanishing boundary terms, the 1st and the 2nd integrals are found to be 
\bea
&&\left.{\rm 1st\ term}=\int_0^{x_{\rm max}}\tau xLdx
=\tau \frac{x^2}{2}L\right\vert_0^{x_{\rm max}}-\int_0^{x_{\rm max}}
\tau\frac{x^2}{2}\frac{d}{dx}\ln\left(\frac{1+\beta}{1-\beta}\right)dx
\nn\\
&&=-\left.\tau x^2\frac{\beta(x)}{1-\beta^2}\right\vert_0^{x_{\rm max}}
+\tau\int_0^{x_{\rm max}}\beta(x)\frac{d}{dx}\left(\frac{x^2}{1-\beta^2}
\right) dx
=-\int_0^{x_{\rm max}}\beta (2x-1)dx
\nn\\
&&=-\frac{\tau}{2(\tau+1)}
+\frac{\tau(\tau+1)\log\left(
\frac{\sqrt{\tau+1}+1}{\sqrt{\tau}}\right)}{2(\tau+1)^{3/2}}~,\\
&&{\rm 2nd\ term}=\int_0^{x_{\rm max}}\beta(2x-1)dx
=\frac{\tau}{2(\tau+1)}
-\frac{\tau(\tau+1)\log\left(
\frac{\sqrt{\tau+1}+1}{\sqrt{\tau}}\right)}{2(\tau+1)^{3/2}}~.
\eea
Therefore the 1st and the 2nd terms cancel with each other and we end up with
\bea
\int_0^{x_{\rm max}} g_1^\gamma(x,Q^2)dx=0~.
\eea

For the quark contribution we have
\bea
&&\int_0^{x_{\rm max}}(2x-1)Ldx
=-\frac{1}{\tau+1}-\frac{\tau\log\left(
\frac{\sqrt{\tau+1}+1}{\sqrt{\tau}}\right)}
{(\tau+1)^{3/2}}~,\\
&&\int_0^{x_{\rm max}}\beta(-4x+3)dx=
\frac{1}{\tau+1}+\frac{\tau\log\left(
\frac{\sqrt{\tau+1}+1}{\sqrt{\tau}}\right)}
{(\tau+1)^{3/2}}~.
\eea
Hence we find 1st+2nd=0. Thus the 1st moment of the
$g_1^\gamma$ structure function for the real photon
target vanishes both for the squark and quark case.

\section{Squark Signature in $W_{TT}^\tau$}
Among the eight virtual photon structure functions, $W_{TT}^\tau$,
which is nothing but a spin-flip helicity amplitude $W(1,1|-1,-1)$,
shows quite different behaviors between squark and quark constituents.
Namely, the squark gives a positive contribution while the
quark contributes negative values for the structure function. 
If we consider the ideal case where the quark and its super-partner
has the same mass, then we have the following relation for the virtual
photon and its real photon limit:
\bea
W_{TT}^\tau \vert_{\rm squark}+W_{TT}^\tau\vert_{\rm quark}=
N_c\frac{\alpha}{\pi}e_q^4\frac{1-\tbeta^2}{\tbeta}L \rightarrow
0 \qquad (P^2 \rightarrow 0)~.
\eea
Namely in the real photon case, both contributions to $W_{TT}^\tau$
exactly cancel each other.
 
\begin{figure}[hbt]
\vspace*{0.3cm}
\begin{center}
\includegraphics[scale=0.7]{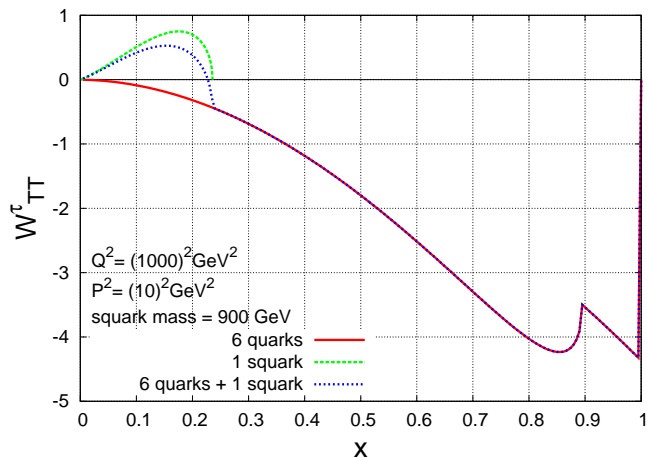}
\caption{\label{WTTtau}
$W_{TT}^\tau$ for a squark with mass 900 GeV (short-dashed curve) and that for 
the six quarks with masses properly taken into account (solid curve)
as well as the total contribution (dotted curve). At around
$x\sim 0.9$ there exists a kink structure due to
the threshold behavior of the top quark. At small $x$ we
find the positive swelling or bump as a signature of the squark contribution.}
\end{center}
\end{figure}

In Fig.\ref{WTTtau} we have plotted the behavior of $W_{TT}^\tau$ for the
six flavor quarks with their masses properly taken into account, as well
as one squark which gives positive contribution. Here we have
taken squark's electric charge to be 2/3 and mass 900 GeV as
an illustration. Note that the signal of the presence of the
squark appears as a positive swelling or bump at small $x$, where 
the quark contributions are negligibly small. 
\footnote{
$W_{TT}^\tau$ can be experimentally measured from the dependence
of the cross section on the azimuthal angle between the scattering 
planes of the electron and positron in the photon center-of-mass 
frame \cite{BGMS}.
}

\section{Conclusion}
In this paper, we have evaluated eight virtual photon structure 
functions arising from squark parton contribution, which have 
been unknown so far. From a general argument based on Cauchy-
Schwarz inequality we can derive the three positivity constraints
on the helicity amplitudes which can then be translated into those
on structure functions.  Now remarkably, for the squark 
contribution, these three constraints turn out to be two equalities 
and one inequality.  

For the case of the real photon target, we obtained the vanishing 
first moment sum rule for the $g_1^\gamma$ structure function, which
is also realized in the case of spin $1/2$ quark parton contribution.
Similarly we have confirmed that the positivity bound
$|g_1^\gamma|\leq F_1^\gamma$ holds for the squark parton as in
the quark parton case. Mass singularities of the structure function
appear in $W_{LT}$ for the case of squark, in contrast to the case of
quark parton where they appear in $W_{TT}$ and $W_{TT}^a$. This can
be interpreted as the spinless nature of squark constituent.

We are particularly interested in the $W_{TT}^\tau$ structure function
for which the behavior of the squark is quite different from that
of the quark. Namely, the squark gives a positive contribution while the
quark contributes negative values for the structure function. 
The signature of the squark could be a positive swelling or bump 
at small $x$, where the quark contributions are negligibly small. 
In the ideal limit where
both squark and quark possess the same mass, their contributions exactly
cancel each other. This situation might be understood from the 
supersymmetric relation.

In our numerical analysis, the kinematic parameters we have chosen for 
the $Q^2$, $P^2$ and $m^2$ are just the illustrative values 
and do not necessarily correspond to the realistic values in 
the ILC region. However, the parameters can be freely scaled up 
or scaled down since we have general formulas for the structure 
functions, which only depend on the ratios such as $m^2/Q^2$, $P^2/Q^2$.
The heavy squark mass, $m$, could be set larger than 1 TeV as the 
recently reported results from the ATLAS/CMS group at LHC.

In this paper we have studied squark contributions to the photon
structure functions only through the QED interaction paying the
particular attention to the heavy mass effects. The logarithmic $Q^2$
dependence due to the supersymmetric QCD radiative effects can be incorporated
by the DGLAP type evolution equation with the suitable boundary condition
taking into account the mass effects \cite{KSUUResum}. 
This will be discussed in the future publication.

\vspace{-0.3cm}
\begin{acknowledgments}
We would like to thank Ken Sasaki for useful discussions about the
evaluation of the photon structure functions. 
Y. K. thanks M. M. Nojiri for the discussion about
the experimental mass bound on SUSY particles.
This work is supported in part by Grant-in-Aid for Scientific Research
(C) from the Japan Society for the Promotion of Science No.22540276.
\end{acknowledgments}


\appendix

\section{Projection Operators} 

In general, taking into account $P$-, $T$-, and gauge invariance, the tensor  $W^{\mu\nu\rho\sigma}$ can be expressed in terms of the eight independent photon structure functions as follows: 
  \begin{eqnarray}
W^{\mu\nu\rho\sigma}&=&(T_{TT})^{\mu\nu\rho\sigma} W_{TT}+(T^{a}_{TT})^{\mu\nu\rho\sigma} W^{a}_{TT}+(T^{\tau}_{TT})^{\mu\nu\rho\sigma} W^{\tau}_{TT}+(T_{LT})^{\mu\nu\rho\sigma} W_{LT}\nonumber\\
&&+(T_{TL})^{\mu\nu\rho\sigma} W_{TL}+(T_{LL})^{\mu\nu\rho\sigma} W_{LL}-(T^{\tau}_{TL})^{\mu\nu\rho\sigma} W^{\tau}_{TL}-(T^{\tau a}_{TL})^{\mu\nu\rho\sigma} W^{\tau a}_{TL}~,
\label{had1}
\end{eqnarray}
where $T_i$'s are the projection operators given by
  \begin{subequations}
    \begin{align}
(T_{TT})^{\mu\nu\rho\sigma}&=R^{\mu\nu}R^{\rho\sigma},\\
(T_{TL})^{\mu\nu\rho\sigma}&=R^{\mu\nu}k_{2}^{\rho}k_{2}^{\sigma},\\
(T_{LT})^{\mu\nu\rho\sigma}&=k_{1}^{\mu}k_{1}^{\nu}R^{\rho\sigma},\\
(T_{LL})^{\mu\nu\rho\sigma}&=k_{1}^{\mu}k_{1}^{\nu}k_{2}^{\rho}k_{2}^{\sigma},\\
(T^{a}_{TT})^{\mu\nu\rho\sigma}&=R^{\mu\rho}R^{\nu\sigma}-R^{\mu\sigma}R^{\nu\rho},\\
(T^{\tau}_{TT})^{\mu\nu\rho\sigma}&=\frac{1}{2}(R^{\mu\rho}R^{\nu\sigma}+R^{\mu\sigma}R^{\nu\rho}-R^{\mu\nu}R^{\rho\sigma}),\\
(T^{\tau}_{TL})^{\mu\nu\rho\sigma}&=R^{\mu\rho}k_{1}^{\nu}k_{2}^{\sigma}+R^{\mu\sigma}k_{1}^{\nu}k_{2}^{\rho}+k_{1}^{\mu}k_{2}^{\rho}R^{\nu\sigma}+k_{1}^{\mu}k_{2}^{\sigma}R^{\nu\rho},\\
(T^{\tau a}_{TL})^{\mu\nu\rho\sigma}&=R^{\mu\rho}k_{1}^{\nu}k_{2}^{\sigma}-R^{\mu\sigma}k_{1}^{\nu}k_{2}^{\rho}+k_{1}^{\mu}k_{2}^{\rho}R^{\nu\sigma}-k_{1}^{\mu}k_{2}^{\sigma}R^{\nu\rho},
    \end{align}
  \end{subequations}
with
  \begin{subequations}
    \begin{align}
R^{\mu\nu}&=-g^{\mu\nu}+\frac{1}{X}\left[p\cdot q(q^{\mu} p^{\nu}+q^{\nu} p^{\mu})-q^2p^{\mu} p^{\nu}-p^2q^{\mu} q^{\nu} \right]~,\\
k_{1}^{\mu}&=\sqrt{\frac{-q^2}{X}}\left(p^{\mu}-\frac{p\cdot q}{q^2}q^{\mu} \right)~,\\
k_{2}^{\mu}&=\sqrt{\frac{-p^2}{X}}\left(q^{\mu}-\frac{p\cdot q}{p^2}p^{\mu} \right)~,
  \end{align}
  \end{subequations}
and
   \begin{equation}
X=(p\cdot q)^2-p^2q^2. 
\end{equation}

\noindent
The unit vectors $k_1$, $k_2$ and the symmetric tensor $R^{\mu\nu}$ which is the metric tensor of the subspace orthogonal to $q$ and $p$, satisfy the following relations:
  \begin{eqnarray}
q\cdot k_1&=&p\cdot k_2=0, \qquad k_{1}^2=k_{2}^2=1~,\nonumber\\
q^{\mu}R_{\mu\nu}&=&p^{\mu}R_{\mu\nu}=k_{1}^{\mu}R_{\mu\nu}=k_{2}^{\mu}R_{\mu\nu}=0, \quad R^{\mu\rho}R_{\rho}^{\nu}=-R_{\mu\nu}~, \label{relationkkR} \\
R_{\mu\nu}R^{\mu\nu}&=&-g^{\mu\nu}R_{\mu\nu}=2~.  \nonumber 
\end{eqnarray}
  
  We also introduce
  \begin{equation}
    x = \frac{Q^2}{2p\cdot q}, \qquad
    Q^2 = -q^2 > 0, \qquad
    P^2 = -p^2 > 0 ,
    \label{eq:080115-08}
  \end{equation}
  and
  \begin{equation}
    \beta=\sqrt{1-\frac{4m^2}{(p+q)^2}}=\sqrt{1+\frac{4m^2x}{xP^2+
(x-1)Q^2}},\quad 
    \tilde{\beta} = \sqrt{1 - \frac{p^2q^2}{(p\cdot q)^2}} = 
\sqrt{1 - \frac{4x^2P^2}{Q^2}} .
    \label{eq:080115-06}
  \end{equation}

  The following relations are useful in the practical calculation:
  \begin{equation}
    \frac{1}{X} = \frac{1}{(p\cdot q)^2 \tilde{\beta}^2} , \qquad
    k_1 \cdot k_2 = \frac{1}{\sqrt{1-\tilde{\beta}^2}} , \qquad
    k_1^\mu k_2^\nu = \frac{\sqrt{1-\tilde{\beta}^2}}{p\cdot q \, \tilde{\beta}^2} \left( p^\mu - \frac{p\cdot q}{q^2} q^\mu \right) \left( q^\nu - \frac{p\cdot q}{p^2} p^\nu \right) ,
  \end{equation}
Unless there is any mass scale in addition to $p^2$, $q^2$ and $p\cdot q$, 
the  structure functions, which are dimensionless,  are
  eventually written in terms of $x$ and $\tilde{\beta}$.  
  
\bigskip  

Using the relations (\ref{relationkkR}), we obtain the following orthogonality and normalization relations:
  \begin{subequations}
    \begin{align}
&      (T_{TT})^{\mu\nu\rho\sigma} (T_{TT})_{\mu\nu\rho\sigma} = 4 ,\qquad
      (T_{TL})^{\mu\nu\rho\sigma} (T_{TL})_{\mu\nu\rho\sigma} = 2 , \nn\\
&      (T_{LT})^{\mu\nu\rho\sigma} (T_{LT})_{\mu\nu\rho\sigma} = 2 , \qquad
      (T_{LL})^{\mu\nu\rho\sigma} (T_{LL})_{\mu\nu\rho\sigma} = 1 , \nn\\
&      (T_{TT}^a)^{\mu\nu\rho\sigma} (T_{TT}^a)_{\mu\nu\rho\sigma} = 4 , \qquad
  (T_{TT}^\tau)^{\mu\nu\rho\sigma} (T_{TT}^\tau)_{\mu\nu\rho\sigma} = 2 , \nn\\
&  (T_{TL}^\tau)^{\mu\nu\rho\sigma} (T_{TL}^\tau)_{\mu\nu\rho\sigma} = 8 , 
\qquad
   (T_{TL}^{\tau a})^{\mu\nu\rho\sigma} (T_{TL}^{\tau a})_{\mu\nu\rho\sigma}= 8~, \nn\\
 &   \hspace{2cm}  (T_i)^{\mu\nu\rho\sigma} (T_j)_{\mu\nu\rho\sigma} = 0 , \qquad \text{for } i \ne j .\nn
    \end{align}
  \end{subequations}
Thus we get the normalized projection operators 
(i.e., $(P_{TT})^{\mu\nu\rho\sigma}W_{\mu\nu\rho\sigma}
=W_{TT}$ and etc.) which  read;
  \begin{subequations}
    \begin{align}
& (P_{TT})^{\mu\nu\rho\sigma} = \frac{1}{4} (T_{TT})^{\mu\nu\rho\sigma} , \qquad  (P_{TL})^{\mu\nu\rho\sigma} = \frac{1}{2} (T_{TL})^{\mu\nu\rho\sigma} , \nn\\
& (P_{LT})^{\mu\nu\rho\sigma} = \frac{1}{2} (T_{LT})^{\mu\nu\rho\sigma} , \qquad  (P_{LL})^{\mu\nu\rho\sigma} = (T_{LL})^{\mu\nu\rho\sigma} , \nn\\
& (P_{TT}^a)^{\mu\nu\rho\sigma} = \frac{1}{4} (T_{TT}^a)^{\mu\nu\rho\sigma} , 
\qquad
  (P_{TT}^\tau)^{\mu\nu\rho\sigma} = \frac{1}{2} (T_{TT}^\tau)^{\mu\nu\rho\sigma} , \nn\\
&  (P_{TL}^\tau)^{\mu\nu\rho\sigma} = - \frac{1}{8} (T_{TL}^\tau)^{\mu\nu\rho\sigma} , \qquad
     (P_{TL}^{\tau a})^{\mu\nu\rho\sigma} = - \frac{1}{8} (T_{TL}^{\tau a})^{\mu\nu\rho\sigma} .\nn
    \end{align}
    \label{eq:080115-02}
  \end{subequations}
\section{The eight virtual photon structure functions}
\bea
&&W_{TT}=N_c\frac{\alpha}{\pi}e_q^4
\left[L\frac{1}{4\tbeta^5}\frac{1}{x}(1-\beta^2\tbeta^2)(4x(1-x)-1+\tbeta^2)
\left\{2x\left(\frac{m^2}{Q^2}-\frac{P^2}{Q^2}(x^2+x-1)\right)\right.\right.
\nn\\
&&\left.\left.
-\frac{1}{2}(1-\tbeta^2)\left[(1-x)(1-\beta^2)
+x\beta^2\frac{P^2}{Q^2}\right]+(2x-1)
\right\}+\frac{\beta}{\tbeta^4}\left\{2x\left(\frac{P^2}{Q^2}(2x(x^2-4x+2)-1)
\right.\right.\right.\nn\\
&&\left.\left.\left.-\frac{2m^2}{Q^2}(x-1)\right)
+\frac{1}{2}(1-\tbeta^2)\left[\frac{m^2}{Q^2}(8x^2-8x-2)+\frac{P^2}{Q^2}
(12x^2-8x+1)\right]\right.\right.\nn\\
&&\left.\left.
+\frac{1}{4}(1-\tbeta^2)^2\frac{1}{x}\left[(1-x)(1-\beta^2)
+x\beta^2\frac{P^2}{Q^2}\right]+(2x^2-2x+1)
\right\}
\right]~,\\
\nn\\
&&W_{TL}=N_c\frac{\alpha}{\pi}e_q^4 2x(1-2x)^2\frac{P^2}{Q^2}\left[
-\frac{1}{2\tbeta^5}L
\left\{2x\left(\frac{P^2}{Q^2}(2x^2-2x+1)-\frac{2m^2}{Q^2}\right)
\right.\right.
\nn\\
&&\left.\left.\hspace{0.7cm}
+(1-\tbeta^2)\left[(1-x)(1-\beta^2)+x\beta^2\frac{P^2}{Q^2}\right]+
2(x-1)\right\}
+3\beta\frac{1}{\tbeta^4}\left(\frac{P^2}{Q^2}x+(x-1)\right)
\right]~,\\
\nn\\
&&W_{LT}=N_c\frac{\alpha}{\pi}e_q^4 
\left(1-\frac{2P^2x}{Q^2}\right)^2
\left[-\frac{1}{\tbeta^5}L
\left\{2x^2\left(\frac{P^2}{Q^2}(2x^2-2x+1)-\frac{2m^2}{Q^2}\right)\right.
\right.\nn\\
&&\left.\left.\hspace{0.3cm}
+(1-\tbeta^2)\left[x(1-x)(1-\beta^2)+x^2\beta^2\frac{P^2}{Q^2}
\right]-2x(1-x)\right\}+\frac{6\beta x}{\tbeta^4}
\left(\frac{P^2}{Q^2}x+(x-1)\right)\right]~,\\
\nn\\
&&W_{LL}=N_c\frac{\alpha}{\pi}e_q^4
\frac{P^2}{Q^2}x
\left[
-\frac{1}{\tbeta^5}L(2x-1)(2x-1+\tbeta^2)\left(
2\frac{P^2}{Q^2}x(2x+3)+6x-7\right)\right.\nn\\
&&\left.\hspace{1cm}+\frac{1}{\tbeta^4}\frac{8\beta x}{1-\beta^2\tbeta^2}
\frac{\frac{P^2}{Q^2}x+(x-1)}{(2x-1)^2-\tbeta^2}
\left\{-2\left[4(x-1)\left(\frac{8m^2x^2}{Q^2}-(1-\tbeta^2)x(2x-3)\right)+
(1-\tbeta^2)\right]\right.\right.\nn\\
&&\left.\left.\hspace{1cm}+2(1-\tbeta^2)x\left[\frac{8m^2}{Q^2}(4x^2-4x-1)+
\frac{P^2}{Q^2}(20x^2-20x+1)\right]\right.\right.\nn\\
&&\left.\left.\hspace{1cm}+4(1-\tbeta^2)^2
\left[(1-x)(1-\beta^2)+x\beta^2\frac{P^2}{Q^2}\right]+2x(1-2x)^2\right\}
\right]~, \\
\nn\\
&&W_{TT}^a=N_c\frac{\alpha}{\pi}e_q^4
\left[\frac{1}{\tbeta^3}L\left\{-(1-\tbeta^2)
\left[1-\beta^2(1-x)+x\beta^2\frac{P^2}{Q^2}\right]
+\frac{4m^2x}{Q^2}+1-\tbeta^2\right\}\right.\nn\\
&&\left.\hspace{3cm}+\frac{\beta}{\tbeta^2}
(2x-1)\left(1-\frac{2P^2}{Q^2}x\right)\right]~,
\eea
\bea
&&W_{TT}^\tau=N_c\frac{\alpha}{\pi}e_q^4
\left[\frac{1}{\tbeta^5}L
\left\{-2x(1-\tbeta^2)\left[1-\beta^2(1-x)+x\beta^2\frac{P^2}{Q^2}
\right]+2(1-\beta^2)x(1-x)\right.\right.
\nn\\
&&\left.\left.\hspace{3cm}
+\frac{1}{2}(1-\tbeta^2)(3+\beta^2)+8x^2-6x(1-\tbeta^2)
\right\}\left\{-\frac{P^2}{Q^2}\left[(1-\beta^2)x(1-x)\right.
\right.\right.\nn\\
&&\left.\left.\left.\hspace{3cm}+\frac{1}{4}\beta^2(1-\tbeta^2)+x^2\right]
+\frac{m^2}{Q^2}+\frac{P^2}{Q^2}x\right\}+
\frac{\beta}{\tbeta^4}\left\{\frac{1}{4}\frac{P^2}{Q^2}(1-\tbeta^2)^2+\frac{1}{2}\frac{P^2}{Q^2}(1-\tbeta^2)
\right.\right.\nn\\
&&\left.\hspace{3cm}\times(20x^2-12x+1)+(1-\tbeta^2)
(x^2-6x+2)+2x^2\right.\nn\\
&&\left.\left.
\hspace{3cm}-\frac{1}{4x}\tbeta^2(1-\beta^2)[4x(1-x)-(1-\tbeta^2)]
\left(\frac{P^2}{Q^2}x+x-1\right)\right\}\right]~,
\\
&&W_{TL}^\tau=N_c\frac{\alpha}{\pi}e_q^4
\sqrt{1-\tbeta^2}\left[
\frac{1}{2\tbeta^5}L\left(4x\left[
\frac{P^2}{Q^2}(2x(x-1)(x-3)-1)-\frac{2m^2}{Q^2}(x-1)\right]
+(1-\tbeta^2)\right.\right.
\nn\\
&&\left.\left.\hspace{1cm}\times\left(
\frac{m^2}{Q^2}(8x^2-8x-2)+\frac{P^2}{Q^2}(8x^2-8x+1)\right)
+2\frac{P^2}{Q^2}x(1-\tbeta^2)\left[(1-x)(1-\beta^2)
+x\beta^2\frac{P^2}{Q^2}\right]
\right.\right.
\nn\\
&&\left.\left.\hspace{1cm}+(1-2x)^2\right)
-\frac{2\beta}{\tbeta^4}\left(x-1+x\frac{P^2}{Q^2}\right)
\left(3x-1-x(8x-3)\frac{P^2}{Q^2}\right)
\right]~,\\
&&W_{TL}^{\tau a}=N_c\frac{\alpha}{\pi}e_q^4
\sqrt{1-\tbeta^2}\nn\\
&&\hspace{2cm}\times
\left[-\frac{1}{2\tbeta^3}L
\left\{(1-\tbeta^2)\left[1-\beta^2(1-x)+x\beta^2\frac{P^2}{Q^2}
\right]
-2x\left(1+\frac{2m^2+P^2}{Q^2}\right)+1\right\}\right.\nn\\
&&\left.\hspace{3cm}
-\frac{\beta}{\tbeta^2}
\left\{(x-1)+x\frac{P^2}{Q^2}\right\}
\right]~.
\eea


\end{document}